\documentclass[10pt, conference, letterpaper]{IEEEtran}
\IEEEoverridecommandlockouts
\usepackage{cite}
\usepackage{amsmath,amssymb,amsfonts}
\usepackage{algorithmic}
\usepackage{graphicx}
\usepackage{textcomp}
\usepackage{xcolor}
\usepackage{epstopdf}
\usepackage{hyperref}
\usepackage{multirow}
\usepackage{subfigure}
\usepackage{enumitem}

\def\BibTeX{{\rm B\kern-.05em{\sc i\kern-.025em b}\kern-.08em
    T\kern-.1667em\lower.7ex\hbox{E}\kern-.125emX}}
\IEEEoverridecommandlockouts \IEEEpubid{\makebox[\columnwidth]{978-3-903176-27-0~\copyright2020 IFIP \hfill} \hspace{\columnsep}\makebox[\columnwidth]{ }}    
    
\begin{document}

\title{Anatomy of Multipath BGP Deployment\\ in a Large ISP Network

\thanks{Jie Li is supported by China Scholarship Council (CSC) with grant no. 201406060022.}
}

\author{\IEEEauthorblockN{Jie Li}
\IEEEauthorblockA{%\textit{dept. name of organization (of Aff.)} \\
\textit{University College London}\\
London, United Kingdom \\
jie.li@cs.ucl.ac.uk}
\and

\IEEEauthorblockN{Vasileios Giotsas}
\IEEEauthorblockA{%\textit{dept. name of organization (of Aff.)} \\
\textit{Lancaster University}\\
Lancaster, United Kingdom \\
v.giotsas@lancaster.ac.uk}
 
\and

\IEEEauthorblockN{Shi Zhou}
\IEEEauthorblockA{%\textit{dept. name of organization (of Aff.)} \\
\textit{University College London}\\
London, United Kingdom\\
s.zhou@ucl.ac.uk}

}

\maketitle

\begin{abstract}
\label{Sec:abstract}
Multipath routing is useful for networks to achieve load sharing among multiple routing paths.
Multipath BGP (M-BGP) is a technique to realize {\em inter-domain} multipath routing by enabling a BGP router to install multiple equally-good routes to a destination {\em prefix}. 
Most of previous works did not distinguish between intra-domain and inter-domain multipath routing. %
In this paper, we present  a measurement study on the deployment of M-BGP in a large Internet service provider (ISP) network. Our method   combines control-plane BGP measurements using Looking Glasses (LG), 
and data-plane traceroute measurements using RIPE Atlas.
We focus on  Hurricane Electric (AS6939) because it is a global ISP that connects with hundreds of major exchange points and exchanges IP traffic with thousands of different networks. And more importantly, we find that this ISP has by far the largest number of M-BGP deployments among autonomous systems with LG servers.
Specifically, Hurricane Electric  has deployed M-BGP with 512 of its peering ASes at 58 PoPs   around the world, including many top ASes and content providers.   
We also observe that most of its M-BGP deployments involve IXP interconnections. 
Our work provides insights into the latest deployment of M-BGP in a major ISP network and it highlights the characteristics and effectiveness of M-BGP as a means to realize load sharing.
\end{abstract}


\section{Introduction}
\label{Sec:Intro}

Multipath routing helps a network obtain higher capacity and performance through load balancing, improve the timeliness of their response to path changes, and enhance their resilience and security in the face of failures and attacks~\cite{Valera2011MBGP}. 
Various approaches have been proposed both to enable multipath routing~\cite{Singh2015IEEECST, Qadir2015IEEECST}, 
and measure the deployment of multipath routes in the Internet~\cite{Augustin2007IMC, Augustin2011TON, Vermeulen2018IMC}. 
However, most of the existing studies~\cite{Augustin2007IMC, Augustin2011TON,Vermeulen2018IMC, Vermeulen2020NSDI,Almeida2020INFOCOM} either focused on intra-domain routing, or did not distinguish between intra-domain  and inter-domain links. 

A key challenge with multipath inter-domain routing is to make the technique compatible with existing BGP semantics and BGP routers~\cite{Valera2011MBGP}. 
Today, most major router vendors, including  Juniper,  Cisco~\cite{CISCO},  and  Huawei, support Multipath BGP (M-BGP)  to enable \textit{load sharing} between inter-domain paths of equal cost. Specifically, when a BGP router learns multiple eBGP paths from the same peering AS to a prefix with equal preference metrics (e.g.~Weight and LocPref), length and MED values, it installs all of these paths together in the routing table instead of trying additional tie-breaking metrics.
Load sharing can be realized per-destination using a hash of the IP headers. M-BGP differs from the other multipath routing techniques in that the multiple equally-good paths are learnt from the same peering AS; and they are for the same destination prefix, not for the same destination IP. 

Aside from limited M-BGP approaches supported by the existing router deployments, the fact that the feature is optional and its application happens only for paths with no tie-breakers in the BGP path selection process, means that its actual deployment and impact on the inter-domain paths is obscure. 
The difficulty in measuring M-BGP paths has been exacerbated by the difficulties in pinpointing the inter-domain borders in traceroute paths. 
Despite over a decade of research in IP-to-AS mapping \cite{Huffaker2010PAM, Pansiot2010PAM, topo-survey-2014, Luckie2016IMC, Marder2016IMC, Marder2018IMC}, accurate border mapping is still a challenge~\cite{yeganeh2019}.
As a result, to the best of our knowledge 
there has not been measurement studies on the deployment of M-BGP. 

In this paper, we present a first step toward this direction by implementing a measurement methodology that combines control-plane BGP measurements using Looking Glasses, 
and data-plane traceroute measurements over RIPE Atlas \cite{RIPE2015IPJ}.
 
We focus on   Hurricane Electric (AS6939) because its LG server provides access to border routers across hundreds of Points-of-Presence (PoPs) where it establishes inter-domain connectivity and it also hosts active RIPE Atlas probes at overlapping locations. 
 In particular, we executed BGP queries to 112 border routers of Hurricane Electric to obtain the peering ASes connected to Hurricane Electric via multiple neighbor addresses.
We then queried a number of /24 prefixes that originate from each of these peering ASes. At last we identified the M-BGP deployment from the responses. In this paper, we only consider prefixes with length of /24 and peering ASes with multiple neighbor addresses via IXPs.
Hence our results provide a lower bound on Hurricane Electric's deployment of M-BGP. 

Our findings reveal a wide deployment of M-BGP in Hurricane Electric. Overall it deploys M-BGP  with 512 peering ASes at 58 PoPs -- more than half of queried PoPs. 
We discover that most of its M-BGP deployments involve IXPs interconnections. 
82.8\% of the M-BGP deployments involve 2 inter-domain, alternative routes, 8.7\%  involve 3 routes, and 8.4\% involve 4 routes. 
We have not observed any deployments involving more than 4 routes.
Those with more than two routes typically involve large Content Provider Networks (CDNs), such as Apple, Cloudflare, or Microsoft.

We then execute a traceroute campaign to study the data-plane behavior of M-BGP load sharing for paths with overlapping locations of RIPE Atlas probes and LG vantage points. 
The traceroute data show that when M-BGP is deployed, the use of multiple inter-domain links is split almost equally between the number of  IPs in the destination prefix. 
The egress link selected for each destination IP remains stable across our measurement period of 4 days indicating that the same per-flow load sharing algorithm is used across all border routers. 
This observation also highlights that M-BGP differs from other multipath routing techniques in that it distributes traffic to IP addresses in the same destination prefix onto different inter-domain links while maintaining a fixed routing path to each IP address.

The techniques and results we present in this paper provide a first step toward developing a more thorough understanding of M-BGP deployment.
We believe that our contributions are relevant to industry stakeholders, Internet engineers and researchers who can apply our techniques to assess the impact of M-BGP on performance, BGP dynamics and the routing behavior under conditions of stress.

\begin{figure*}[!t]
    \centering
     \includegraphics[width= 1 \textwidth]{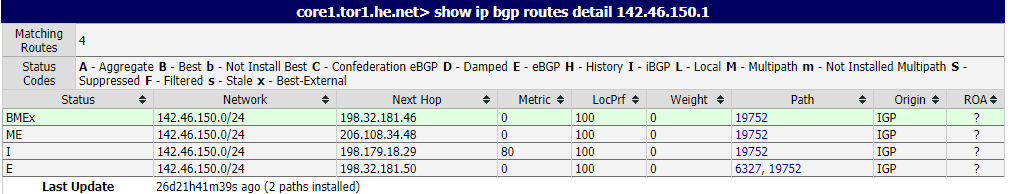}
    \caption{Example of LG response to the command of {\tt show ip bgp route detail}}
    \label{fig:Responses}
\end{figure*}

\section{Multipath Routing}
\label{Sec:MultipathRouting}

Detection of different IP-level routing paths between a pair of hosts  has been the basis for the study of `anomalies' and `routing dynamics~\cite{Ahmed2015LCN}'. There was a considerable effort to characterise~\cite{Medem2012INFOCOM, Comarela2013IMC, Rimondini2014PAM, Fanou2015PAM} and predict changing patterns of routing paths~\cite{Cunha2014TON, Wassermann2017BigDAMA}. 
While  some of the observed different routing paths  could indeed be due to  anomalies or routing dynamics, it is now understood that many of them could  be legitimate  routes due to multipath routing~\cite{Nur2018Comput.Netw., Mok2018PAM}.
Indeed, in recent years network operators and service providers increasingly utilize multipath routing  for traffic load balancing and load sharing to improve performance and resilience~\cite{Singh2015IEEECST}.
Multipath routing has attracted significant research attention, with proposals that span different layers, protocols, and techniques~\cite{Qadir2015IEEECST}. 

Augustin \textit{et al.}~\cite{Augustin2011TON} presented one of the first measurement studies of multipath routing by developing the Multipath Detection Algorithm (MDA) to  identify diamond-shaped IP-level routing paths in traceroute data. 
Their work focused mostly within the boundary of a domain and as they remarked ``\textit{the traditional concept of a single  network path between hosts no longer holds}''. 
MDA was first proposed in \cite{Augustin2007E2EMON} to detect multipath routing from a single source and a single destination. It used Paris traceroute and adapted the number of probes to send hop by hop, in order to find as many load balancing behaviors as possible. It was then improved with a number of follow-up modifications \cite{Veitch2009INFOCOM, Vermeulen2018IMC}.

Recently, a number of research works have extended MDA to improve the completeness of load balancing identification in traceroute paths and reduce the measurement cost. 
Vermeulen \textit{et al.}~\cite{Vermeulen2020NSDI} introduced D-Miner to discover load-balanced paths at scale by utilizing the high-speed probing techniques of Yarrp~\cite{yarrp}.
Almeida \textit{et al.}~\cite{Almeida2020INFOCOM} proposed Multipath Classification Algorithm (MCA) to identify and classify load balancing in the Internet. Specifically, it extended the existing formalism and router model of  \cite{Veitch2009INFOCOM}, and the discovery techniques of \cite{Augustin2011TON} to capture paths that relied on arbitrary per-packet load balancing.

In addition to the above works based on active traceroute probing, researches also studied multipath routing and routing diversity based on existing traceroute datasets. 
Vanaubel \textit{et al.} \cite{Vanaubel2015IMC} proposed the Label Pattern Recognition algorithm for analyzing traceroute data (from CAIDA Archipelago \cite{Ark}) with Multiprotocol Label Switching (MPLS) information, which provided insights into transit path diversity in an ISP's network.
Iodice \textit{et al.}~\cite{Iodice2019IEEEAccess} reported that a  large percentage of  traceroutes from RIPE Atlas exhibit a periodic behavior, where a small amount of path changes were related to  MPLS and load balancing.

The multipath routing or load balancing behavior studied in the above-mentioned studies
did not distinguish intra-domain from inter-domain links. Our work studies the load balancing on inter-domain links and in particular the deployment of Multipath-BGP, which differs from these multipath routing in  installing multiple equally good paths to the same destination prefix learnt from the same peering AS.

In terms of inter-domain routing, Giotsas~\textit{et al.} introduced the Constrained Facility Search algorithm, which used topology data from different levels of abstractions to map IP connectivity to PoPs~\cite{Giotsas2015CoNEXT}.
Motamedi~\textit{et al.}~\cite{mi2} presented the $mi^2$ (mapping Internet interconnections) algorithm that improved PoP mapping through more accurate identification of inter-domain borders.
Nur and Tozal~\cite{Nur2018Comput.Netw.} presented the cross-AS topology maps and defined the cross-border interfaces to study relevant topological properties. 
However, the above works focused on Internet mapping or topological properties, lacking of knowledge on how the diverse inter-domain connectivity was used in multipath routing.
The closest work to ours is by Mok \textit{et al.}~\cite{Mok2018PAM}, which studied the load-balancing behavior on inter-domain links by YouTube with data-plane data, i.e., traceroute data. Our work focuses on studying the deployment of M-BGP with control-plane data provided by LG server  and data-plane measurements on RIPE Atlas.

\section{Multipath BGP (M-BGP)}
\label{Sec:M-BGP}

By default, BGP requires that for each prefix a single ``best'' path should be installed in the routing table to be used for traffic forwarding and be advertised to the BGP sessions~\cite{RFC4271}.
To rank all the available paths BGP uses a multi-step decision process that examines a series of attributes in strict order.
While the actual metrics may differ across different vendors, almost all major deployments consider the Local Preference, the AS path length and the Multi-Exit Discriminator (MED) values as part of their path selection process.
Local Preference ({\tt LocPref}) is a numerical value that can be set arbitrarily for each path to denote the preference of a route. 
The path with the highest {\tt LocPref} value will be selected as the most  preferable and will be installed in the routing table. 
{\tt LocPref} is assigned locally to a router and is not propagated through BGP updates.
If two routes have equal {\tt LocPref} values, the path with the shortest path, namely the smallest number of AS hops, will be preferred.  
If both {\tt LocPref} and path length cannot determine the best path, BGP continues the path selection process by checking the protocol through which a value is received, and then prefers paths with the lowest MED value if they are received from the same AS neighbor.

Although not defined in the original standard, most major router vendors, including Juniper, Cisco \cite{CISCO}, and Huawei, have added optional support for multipath BGP in the case of Equal Cost Multipath Routing (ECMP). 
If multipath BGP is activated, when there are multiple equally good eBGP paths learnt from the \textit{same peering AS}, and all the first six attributes of the BGP decision process (i.e., {\tt LocPref}, AS Path, Origin, MED, eBGP/iBGP, and Metric) have the same value, instead of comparing the Router ID as a last-resort tie-breaker, multipath BGP allows the router to install more than one paths learnt from different border routers. 
The {\tt maximum-paths} configuration controls the number of paths to be used. 

Load sharing can then happen per-destination using a hash of the IP headers, or per-packet using balanced or weighted round robin~\cite{cisco-book}. While per-packet load balancers have been found to be less frequent~\cite{Augustin2011TON, Vermeulen2020NSDI, Almeida2020INFOCOM}, their deployment may have been underestimated in the past~\cite{Almeida2020INFOCOM}. 
The default M-BGP deployment uses a per-destination hash function, therefore M-BGP provides per-flow load sharing among different IP destinations in the same IP prefix. 
The amount of traffic or the available link capacities are not considered in the default load sharing functionality of M-BGP. 
Nonetheless, operators are able to override the default M-BGP behavior and implement either weighted load-sharing to reflect link capacities, or per-packet load balancing. 

The studies on M-BGP are limited in literature. 
Valera \textit{et al.}~\cite{Valera2011MBGP} explained the motivations to apply M-BGP and discussed some alternatives to M-BGP for achieving multipath routing.
Therefore, while M-BGP is the de-facto technique to achieve load balancing between ASes, we lack insights with regards to the level of its deployment in the Internet.

\section{Looking Glass Announcements of M-BGP}
\label{Sec:LG}

Despite the extensive work in the enumeration of multipath routes in traceroute paths, distinguishing inter-domain from intra-domain multipath routing can be particularly challenging due to the difficulties in identifying the border routers between ASes. 
Traditionally, IP-to-AS mapping has been used to detect pairs of consecutive IP hops that belong in different ASes, and infer the AS border at these IPs. In recent years a number of border mapping techniques have found that such border identification can lead to inaccurate mapping since neighboring ASes may number their interfaces with IPs of neighboring ASes~\cite{Luckie2016IMC,Marder2016IMC}. Accordingly, novel border mapping techniques have been introduced to address these issues with bdrmapIT~\cite{Marder2018IMC} considered as the state-of-the-art.
However, recent works have found that even bdrmapIT can lead to erroneous border identification~\cite{yeganeh2019}.
Therefore, identifying M-BGP through traceroutes alone can lead to a non-trivial amount of false-positives.

To alleviate this issue, we utilize Looking Glasses which can provide a direct and reliable source of information on M-BGP deployment, since they allow to query directly the BGP configuration and routing table of border routers, and obtain BGP information beyond what is propagated through BGP updates in RouteViews~\cite{RouteViews} and RIPE RIS collectors.

\subsection{Looking Glass (LG) servers}
\label{Sec:LG-LGserver}

Many network operators host LG servers, which provide Web-based interfaces to allow non-privileged execution of network commands (e.g., traceroute, ping, and BGP) at one or more border routers for  network measurement and diagnosis~\cite{Khan2013IMC}. LG servers enable researchers and network operators to study  a network's performance from the perspective within the network. 
Different LG servers may provide different sets of commands.
LG routing data, along with other  data sources like RouteViews \cite{RouteViews}, have been widely used in studies on the Internet topology and path diversity~\cite{Chang2004Compt.Netw.,Zhang2005CCR,Han2006TDS,Khan2013IMC}. 
More recently the Periscope platform was proposed~\cite{Giotsas2016PAM} to unify LG servers with publicly accessible querying API and to support on-demand measurements. 

In January 2020, more than 1,200 ASes have LG servers distributed across the world, including many top-ranked ASes \cite{BGPLGDatabase, PeeringDBAPI}.

\subsection{Identifying M-BGP in LG announcement}
\label{Sec:LG-M-BGP}

Some LG servers provide information on whether and how an AS deploys M-BGP with its peering ASes  in   responses to the command  {\tt show ip bgp detail <IP address>}. 

Figure~\ref{fig:Responses} shows an example response from {\tt tor1}, a border router of Hurricane Electric.
There are  two different routes (the two `Next Hops' 198.32.181.46 and 206.108.34.48) towards the same destination prefix (142.46.150.0/24 in Hydro One). They are labelled with status codes of ``M'' and ``E'', meaning these are multipath routes (M) learnt via external BGP (E). Both routes have the same values for all routing metrics, including LocPref, Weight and Path.
Such LG response provides the ground-truth that Hurricane Electric deploys M-BGP with the peering AS Hydro One at the border router {\tt tor1}.

\begin{table}[!t]
    \centering
        \caption{Geographical distribution of border routers\\of Hurricane Electric.}
    \label{tab:BRDistribution}
    \begin{tabular}{l|c c}
    
    \hline
      &   Number of  & with M-BGP \\  
    & border routers  &   deployment\\ \hline

    \textbf{North America} & \textbf{55} & \textbf{24} \\ 
        {~~~~United States}  & 47 & 19 \\
        {~~~~Canada}  & 8 & 5 \\\hline
        
    \textbf{Europe } & \textbf{40} & \textbf{26} \\ 
        {~~~~Germany} & 5 & 4 \\ 
        {~~~~United Kingdom} & 3 & 2 \\
        {~~~~France} & 2 & 2 \\
        {~~~~Other} & 30 & 18 \\ \hline
        
    \textbf{Asia} & \textbf{6} & \textbf{4} \\\hline
    \textbf{Other} & \textbf{11} & \textbf{4} \\\hline
    \textbf{~~~~~~~~Total} & \textbf{112} & \textbf{58} \\ \hline
    
    \end{tabular}

\end{table}

\section{Case Study on Hurricane Electric (AS6939)}
\label{Sec:HE}

To reveal more details on the M-BGP deployment, we conducted a thorough analysis of the Hurricane Electric connectivity.

Hurricane Electric's LG \url{lg.he.net} covers border routers in 112 PoPs. As shown in Table \ref{tab:BRDistribution}, these PoPs are located in 43 countries around the world. They support ping, traceroute, BGP route, BGP summary (IPv4) and BGP summary (IPv6). As a first step, we only study M-BGP on IPv4.  
 
\subsection{Identifying M-BGP}
\label{Sec:HE-M-BGP}

The LG command of {\tt show ip bgp detail <IP address>} requires a target IP address as a parameter, so we need to compile a list of targets.

\begin{figure}
    \centering
    
    \includegraphics[width = 0.48\textwidth]{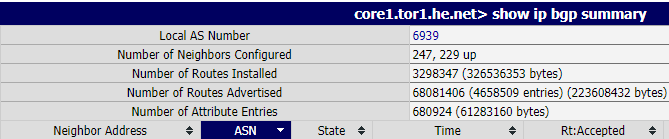}
    \includegraphics[width = 0.48\textwidth]{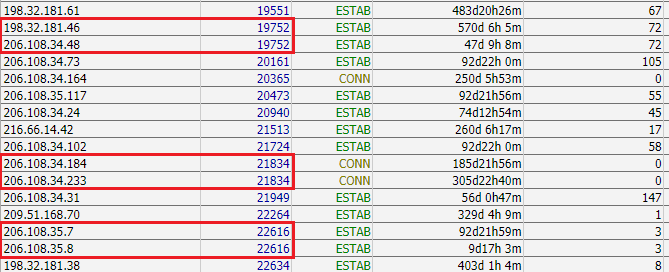}
  
    \caption{Example of LG response to the command {\tt show ip bgp summary}. Each red rectangle  highlights an example of a peering AS with multiple neighbor addresses.}
    \label{fig:AS6939BGPSummary}
\end{figure}

We first query each of Hurricane Electric's border routers with the command {\tt show ip bgp summary} to obtain the BGP connectivity of Hurricane Electric at each of the corresponding locations. 
The command returns a summary table with the ASNs of the BGP neighbors and the addresses of the remote IP interfaces through which the BGP session is established. 
Figure \ref{fig:AS6939BGPSummary} is an example table  from the border router {\tt tor1}, listing the information about each peering AS at this router. 
We find out whether a peering AS is connected via IXP by cross-checking a neighbor address and the peering AS using PeeringDB \cite{PeeringDB} data.

\begin{figure}
    \centering
    {\includegraphics[width = 0.5\textwidth]{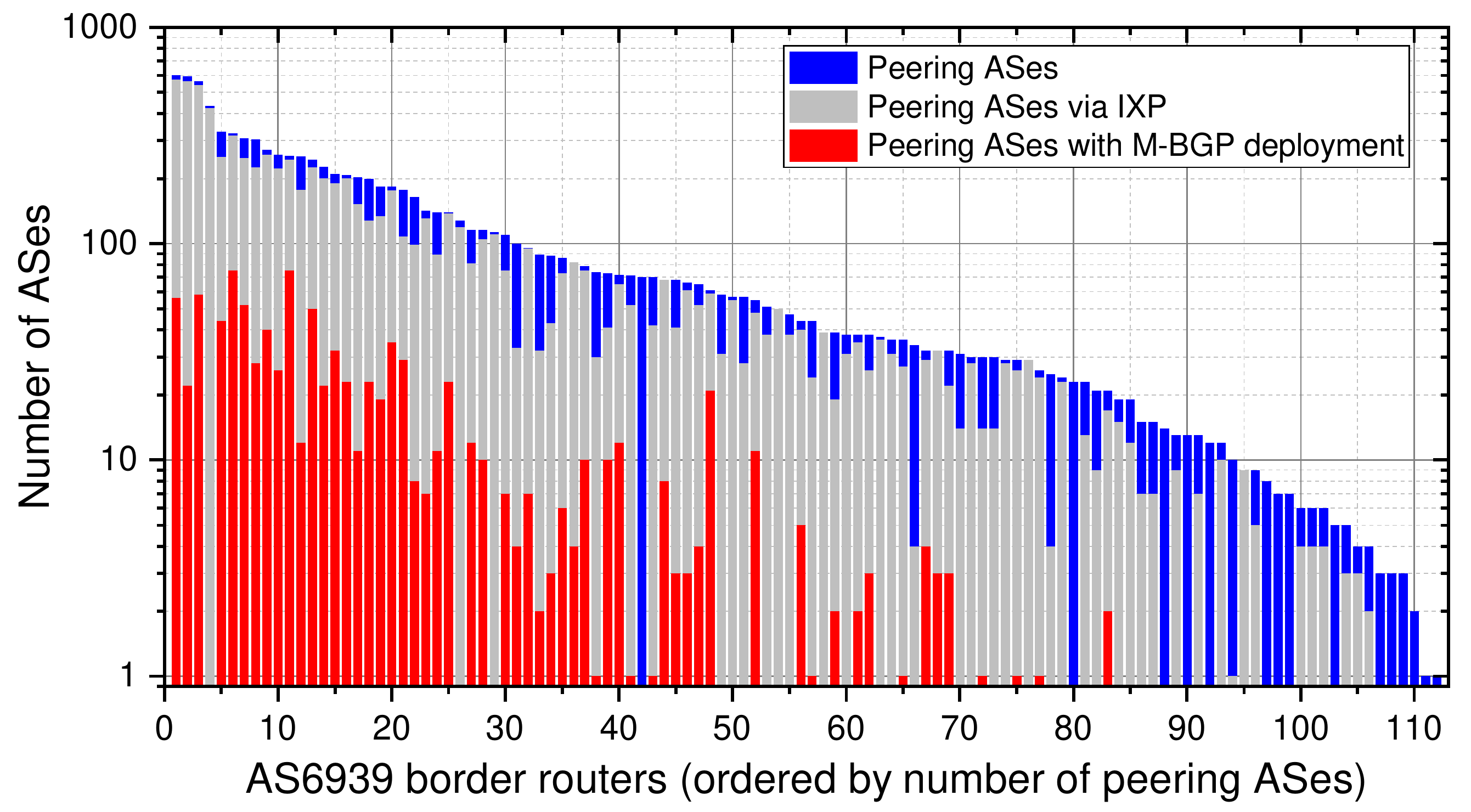}}
    \caption{List of 112 border routers of Hurricane Electric (AS6939). The border routers are ordered by the number of peering ASes at each router. Most of the ASes are peered via IXP. All M-BGP deployments involve IXP.   }
    \label{fig:BR_ASes}
\end{figure}

Figure \ref{fig:BR_ASes} shows the  number of peering ASes connected at each of the 112 border routers of Hurricane Electric.  In total, Hurricane Electric is peering with 5,868 unique ASes, of which 4,622 ASes are peered at 97 border routers via IXPs. 
This result highlights the role of IXPs in providing interconnection between AS peers. 
Note that a peering AS may be counted by a set of  border routers. 

Then, we search for those peering ASes with multiple Neighbor Addresses in the bgp summary (such as AS19752, AS21834 and AS22616 highlighted in Figure 2),  which means Hurricane Electric has multiple inter-domain border links to each of these peering ASes and this is the condition for multiple paths to be {\em tied} before M-BGP is installed. 

For each of these peering ASes, we obtain the list of prefixes from BGP announcement provided by RouteViews \cite{RouteViews} in March 2020. For simplicity, we only study /24 prefixes because (1) /24 prefixes are the most common prefixes installed in BGP routing, e.g. around 60\% of prefixes in the RouteViews data are /24 prefixes; and (2) more importantly, our purpose is to find any evidence of M-BGP deployment with a peering AS, where any of the peering AS’s prefixes can provide sufficient evidence, regardless of its size.

Finally, we query each of Hurricane Electric's border routers using command {\tt show ip bgp detail <IP address>}, where {\tt IP address} is set as {\tt x.x.x.1} for each of the obtained destination prefixes of a peering AS. From each response, we identify whether M-BGP is deployed with the peering AS at the border router towards the destination prefix as explained in Section~\ref{Sec:LG-M-BGP}. 
Note that peering ASes announce different numbers of prefixes. For each peering AS, our query stops as soon as any of its prefix is identified as having M-BGP, because M-BGP should be activated for every prefix learnt through the same set of neighbor interfaces.

\subsection{Results}
\label{Sec:HE-Results}

Querying the LG server is time-consuming because we should avoid violating the querying rate limitation set by Hurricane Electric. By the time we write this paper, we have identified 950 cases of M-BGP deployment by Hurricane Electric with 512 (around 9\% in 5,868) peering ASes at 58 border routers. 
Figure \ref{fig:BR_ASes} plots in red the number of peering ASes with M-BGP deployment at each border router. Note that a peering AS may be deployed with M-BGP for different prefixes at a set of border routers.
Table~\ref{tab:BRDistribution} shows the 58 border routers with M-BGP deployment are distributed around the world.  

\begin{figure}
    \centering
    \includegraphics[width = 0.48 \textwidth]{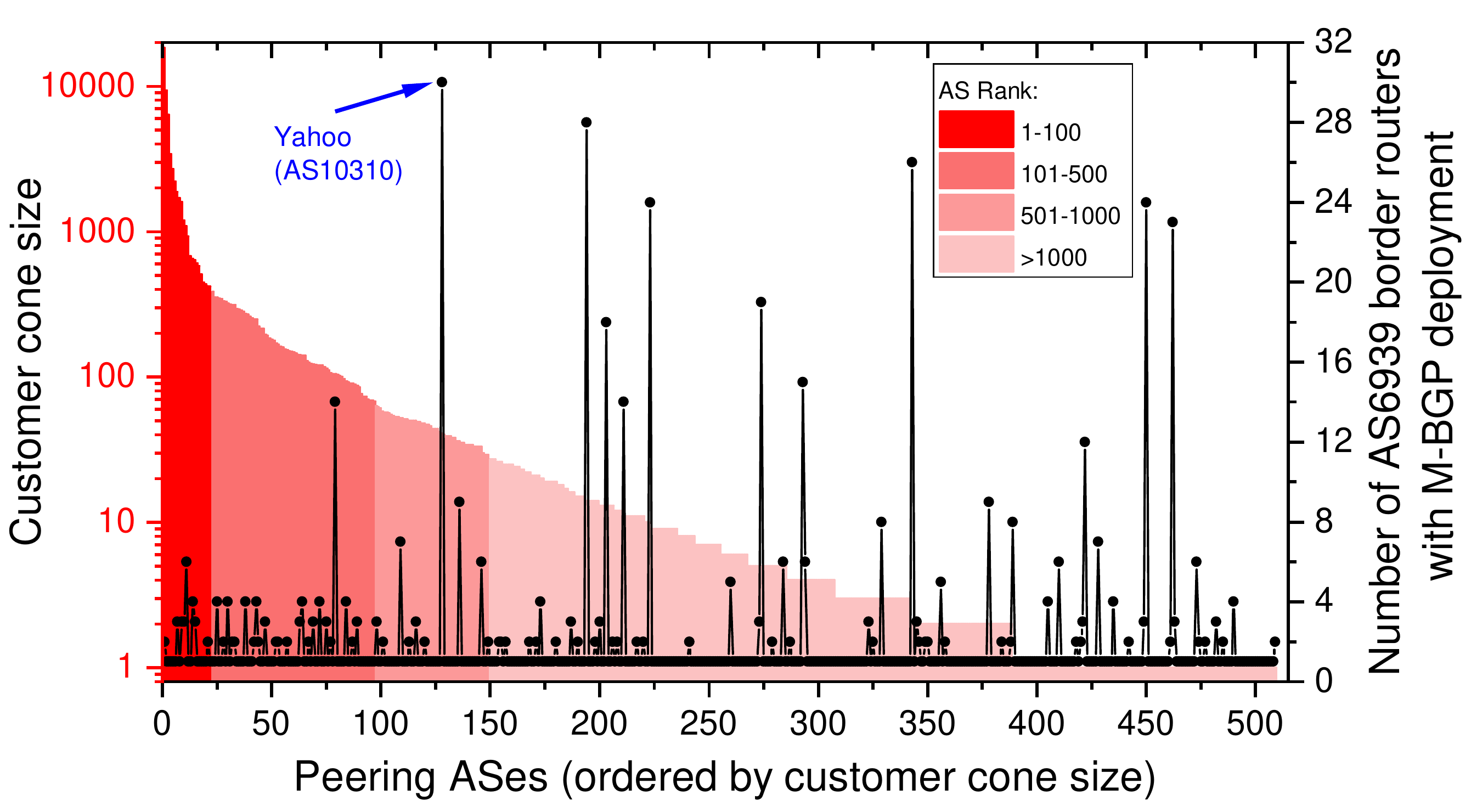}
    \caption{List of 509 peering ASes of Hurricane Electric (AS6939). The ASes are ordered by their customer cone size. Also shown is the number of AS6939's border routers with M-BGP deployment for each peering AS. For example, Yahoo (AS10310) has a customer cone size of 41; its CAIDA AS rank is 754; and Hurricane Electric deploys M-BGP with Yahoo at 30 border routers.}
    \label{fig:peeringASes}
\end{figure}

Figure \ref{fig:peeringASes} plots 509 of the peering ASes with M-BGP deployment, ranked by the size of their customer cone \cite{Luckie2013IMC} (in red). 
The customer cone of an AS is the set of all ASes that the AS can reach via customer links, including customers of its customers, recursively. The size of customer cone can be used as a measure of an AS's influence \cite{Luckie2013IMC} and is used by CAIDA to rank ASes \cite{caida_asrank}. 
These ASes are in four groups by their AS ranks in CAIDA's AS Rank data \cite{caida_asrank}, with the numbers of ASes in each group 22, 75, 52, and 360, suggesting Hurricane Electric deploys M-BGP widely with its peering ASes among different rank groups. Note that 3 ($= 512 - 509$) ASes are missing in the plot because the data snapshot \cite{caida_asrank} we use does not provide the information for them. The plot also shows in black the number of border routers where each peering AS is deployed with M-BGP. We can observe from the figure that the low rank ASes are more likely to be deployed with M-BGP at multiple border routers, suggesting Hurricane Electric's richer connection to low-rank ASes than to top-rank ASes. Table \ref{tab:10ASes} lists the 10 highest ranked peering ASes with M-BGP deployment. 

\begin{table}[t]
    \centering
    \caption{The 10 highest ranked peering ASes with M-BGP deployed at border routers of Hurricane Electric (AS6939).  
    }
    \label{tab:10ASes}
    \begin{tabular}{c|c|r|c|c}
    \hline

      CAIDA   &    & Customer   &  & \# of AS6939\\
       AS  & AS &  cone &  &  border routers  \\ 
     Rank & Number  &   size & AS Name &  with M-BGP\\ \hline
     
       8  & 13101 & 18,372  & ennit server GmbH              & 2 \\ \hline
       12 & 6461  &  9,368  & Zayo Bandwidth                 & 1 \\ \hline
       16 & 9002  &  6,366  & RETN Limited                   & 1 \\ \hline
       22 & 12389 &  3,415  & PJSC Rostelecom                & 1 \\ \hline
       28 & 3216  &  2,691  & PJSC ``Vimpelcom''             & 1 \\ \hline
       33 & 6830  &  2,204  & Liberty Global B.V.            & 1 \\ \hline
       40 & 8359  &  1,867  & MTS PJSC                       & 3 \\ \hline
       41 & 286   &  1,705  & KPN B. V.                      & 1 \\ \hline
       42 & 58453 &  1,601  & China Mobile                   & 3 \\ \hline
       51 & 41095 &  1,198  & IPTP LTD                       & 3 \\ \hline
       
    \end{tabular}
\end{table}

\begin{figure}[!t]
    \centering
    \subfigure[Number of peering ASes  as a function of the number of border routers.]{
    \label{fig:BRandAS-a}
    \includegraphics[width = 0.48\textwidth]{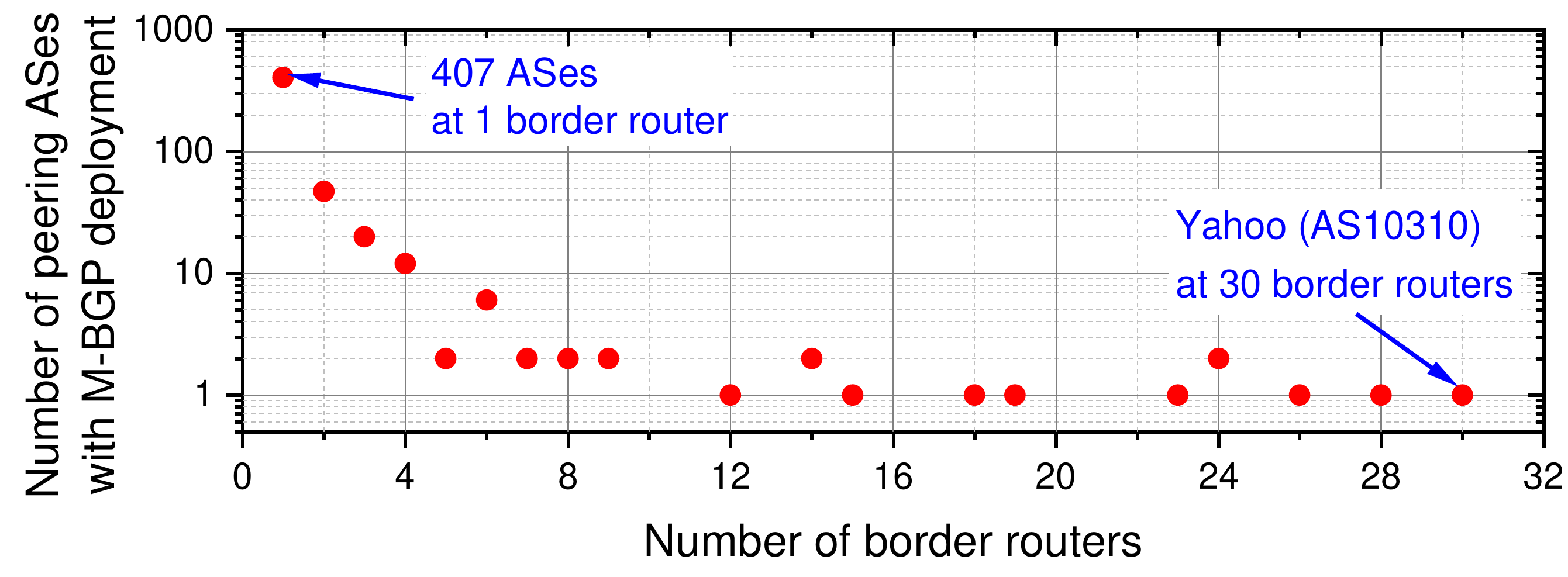}
    }
    \subfigure[Number of border routers as a function of the number of peering ASes.]{
    \label{fig:BRandAS-b}
    \includegraphics[width = 0.48\textwidth]{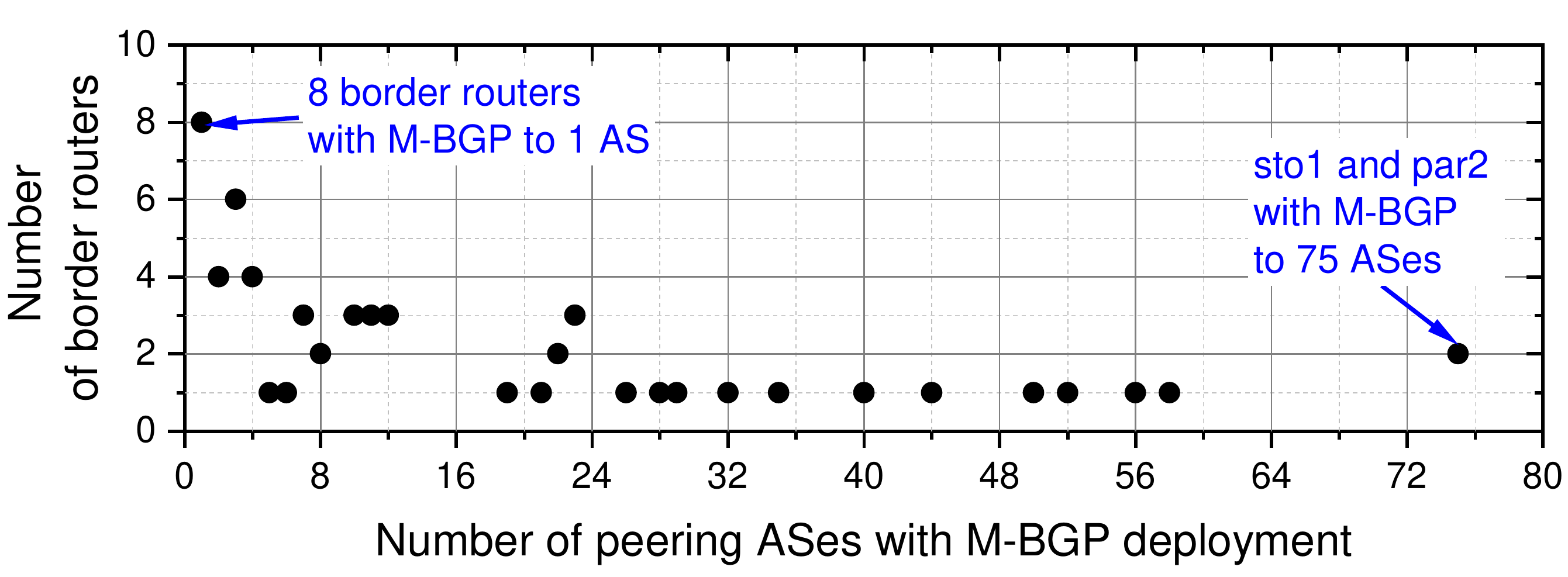}
    }
    \caption{Relation between the number of border routers and the number of peering ASes with M-BGP deployment in Hurricane Electric.}
    \label{fig:BRandAS}
\end{figure}

Figure \ref{fig:BRandAS} shows the relation between the number of border routers and the number of peering ASes with M-BGP deployment. Figure \ref{fig:BRandAS-a} shows that 407 peering ASes are deployed with M-BGP at one border router, and 105 ($=512-407$) peering ASes are deployed with M-BGP at multiple border routers.  Among the peering ASes, Yahoo (AS10310) is deployed with M-BGP at the largest number of  border routers (30), which is also labelled in Figure \ref{fig:peeringASes}. Figure \ref{fig:BRandAS-b} shows that Hurricane Electric  at 8 border routers with only one peering AS, while it deploys M-BGP with multiple peering ASes at 50 ($=58-8$) border routers. Among the border routers, {\tt sto1} and {\tt par2} are both deployed M-BGP to the  most (75) peering ASes. 

Among the 950 cases of M-BGP deployment, 787 (82.8\%) cases are with 2 inter-domain links, 83 (8.7\%) cases are with 3 inter-domain and 80 (8.4\%) cases are with 4 inter-domain links. Moreover, M-BGP paths with more than 2 inter-domain links are predominantly through large CDNs who have elevated capacity requirements. Our results confirm previous studies that found that the so-called Internet hyper-giants rely increasingly on IXPs as part of their content delivery backbone~\cite{netflix, hyper-giants}.

In summary, our result suggests that Hurricane Electric has deployed M-BGP widely, with around 9\% of its peering ASes  at more than half of its border routers distributed  around the world. We confirm the vital role IXPs play in Hurricane Electric's peering fabric and deployment of M-BGP.
Note that we only consider prefixes of length /24 provided by RouteViews and peering ASes with multiple neighbor addresses via IXP. Thus, our result provides a lower  bound of Hurricane Electric's M-BGP deployment.

\begin{table*}[]
    \centering
    \caption{Two types of M-BGP deployed in Hurricane Electric.  For each example case, we run traceroute measurements from a RIPE Atlas probe in Hurricane Electric (source)  to each IP in the destination prefix in a neirghbor AS. The traceroute reveals the nearside IPs (ingress interfaces of nearside border router), IXP IPs, farside IPs and the percentage of routes (\%)  to the IP addresses allocated on each of border links. Figs. \ref{fig:parallel} and \ref{fig:divergent} illustrate the  topology and routing for Cases 1 and 3.}
    \label{tab:cases}
    \begin{tabular}{c|c||c|c|c||c|r||r|c }
    
    \hline\hline
 {M-BGP} &  Case    & Traceroute & Border  &\multirow{2}{*}{Nearside IPs } & \multirow{2}{*}{IXP name}& \multirow{2}{*}{IXP IPs: routes\% }&  \multirow{2}{*}{Farside IPs: routes\%} & {Destination}\\
 {Type}   & No. & source   & router &  &   &  &  & ASN \& prefix\\   
      \hline  \hline

% ----Case 1
 & \multirow{2}{*}{1} & \multirow{2}{*}{65.49.77.70} &  \multirow{2}{*}{{\tt sjc2}}	& {184.105.213.157} &  Equinix & 	{206.223.117.58: 50.0\%} &	{X: 199.230.0.190: 50.0\%} &  AS14630 \\ \cline{5-5} \cline{7-8}
\multirow{2}{*}{Parallel} & & 	    & & {72.52.92.246} & San Jose	& {206.223.117.57: 50.0\%} &	{Y: 199.230.0.182: 50.0\%} &  142.148.224.0/24  \\  \cline{2-9}

% ----Case 2
 & \multirow{2}{*}{2}& 	\multirow{2}{*}{65.49.77.70}  & \multirow{2}{*}{{\tt sjc2}}	& {184.105.213.157}  & Equinix &	{206.223.117.18: 50.0\%} &	{64.16.254.8: 48.4\%} &AS63440 \\ \cline{5-5} \cline{7-8}
 & &	  &  & {72.52.92.246} & San Jose &	{206.223.116.110: 50.0\%} &	{64.16.254.2: 50.0\%}  &  192.76.120.0/24	 \\ 
 \hline \hline
 
% ----Case 3
 & &  & & &  	& &	A: 74.122.191.5~: 19.5\% &	\\ \cline{8-8}
 &  & &	 & 184.105.213.157 &	&	206.223.116.50: 49.5\% & B: 74.122.191.25: 12.6\% &	 \\ \cline{8-8}
 & \multirow{2}{*}{3}  & \multirow{2}{*}{65.49.77.70}	 & \multirow{2}{*}{{\tt sjc2}} & & Equinix &	 &	C: 74.122.191.35: 17.4\%  &  AS15211\\ \cline{5-5} \cline{7-8}
 &  &	 &	 & & San Jose	 &	& D: 74.122.191.7~: 18.2\% & 74.122.186.0/24	 \\ \cline{8-8}
 \multirow{2}{*}{Divergent} &  &	  &	 & 72.52.92.246 &	&	206.223.116.49: 50.5\% & E: 74.122.191.27: 11.0\% &	 \\ \cline{8-8}
 & & 		&  &	 &  &	 &	F: 74.122.191.37: 21.3\% &   \\  \cline{2-9} 

% ----Case 4
 & &	  & &	 &Equinix &	\multirow{2}{*}{198.32.181.46: 50.3\%} &	142.47.202.50: 25.1\% &\\ \cline{8-8}
 &\multirow{2}{*}{4} & \multirow{2}{*}{209.51.186.5}	  & \multirow{2}{*}{{\tt tor1}}&		\multirow{2}{*}{209.51.161.49} & Toronto & &	142.47.203.14: 25.2\% & AS19752 \\ \cline{6-8}
 & & 	   & &	 & \multirow{2}{*}{TorIX} &	\multirow{2}{*}{206.108.34.48: 49.7\%} &	142.47.202.50: 24.0\% & 142.46.150.0/24 \\ \cline{8-8}
 & &	 & &	& &	 &	142.47.203.14: 25.7\% & \\ \hline\hline

    \end{tabular}
\end{table*}
\begin{figure*}
    \centering
    \subfigure[Topology]{
     \label{fig:topology_case1}
    \includegraphics[width = 0.60 \textwidth]{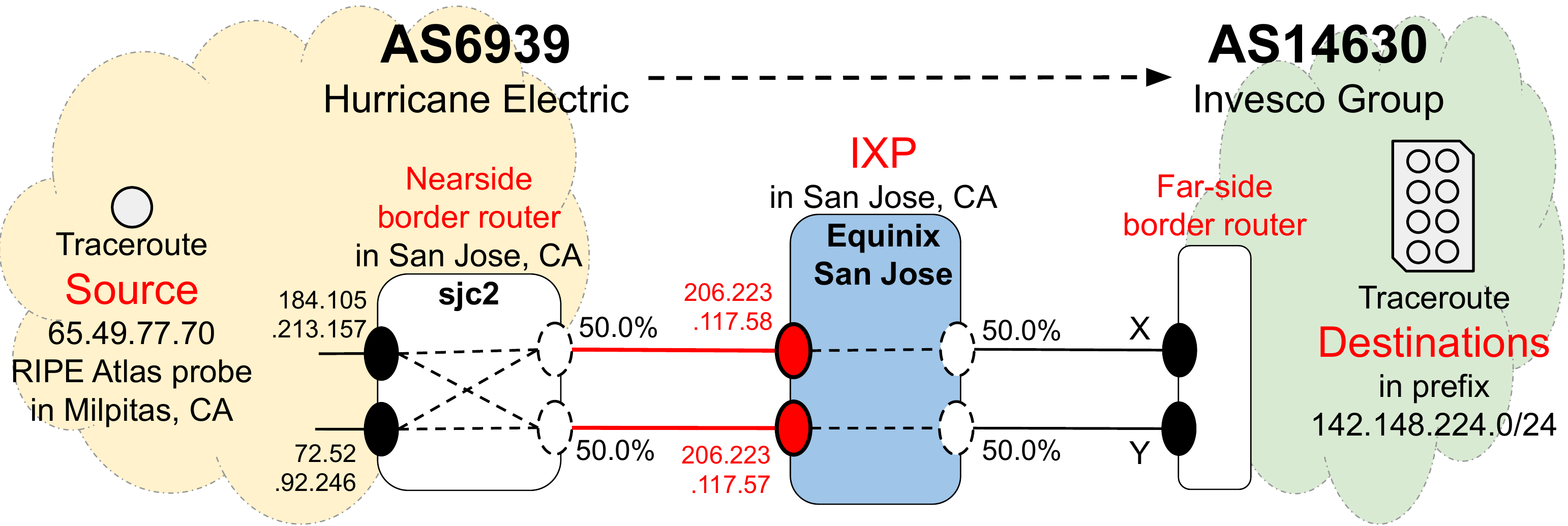}
    }
    \subfigure[Routing]{
     \label{fig:traffic_case1}
    \includegraphics[width = 0.36 \textwidth]{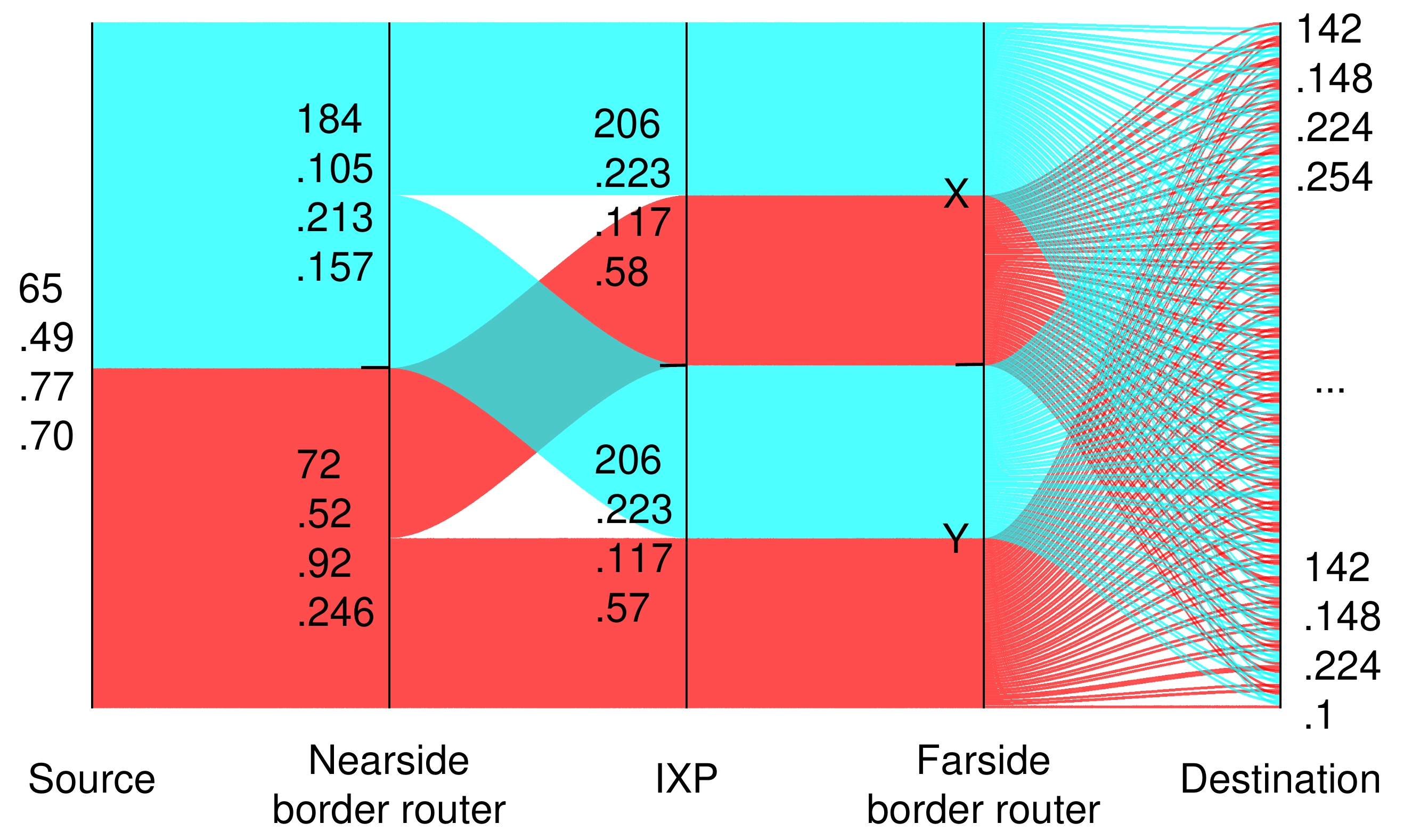}
    }
    \caption{Illustrations of topology and routing of a Parallel-type M-BGP deployment between Hurricane Electric and Invesco Group (Case 1 in Table \ref{tab:cases}). 
    }   
    \label{fig:parallel}
\end{figure*}
\begin{figure*}
    \centering
   \subfigure[Topology]{
   \label{fig:topology_case3}
    \includegraphics[width = 0.60 \textwidth]{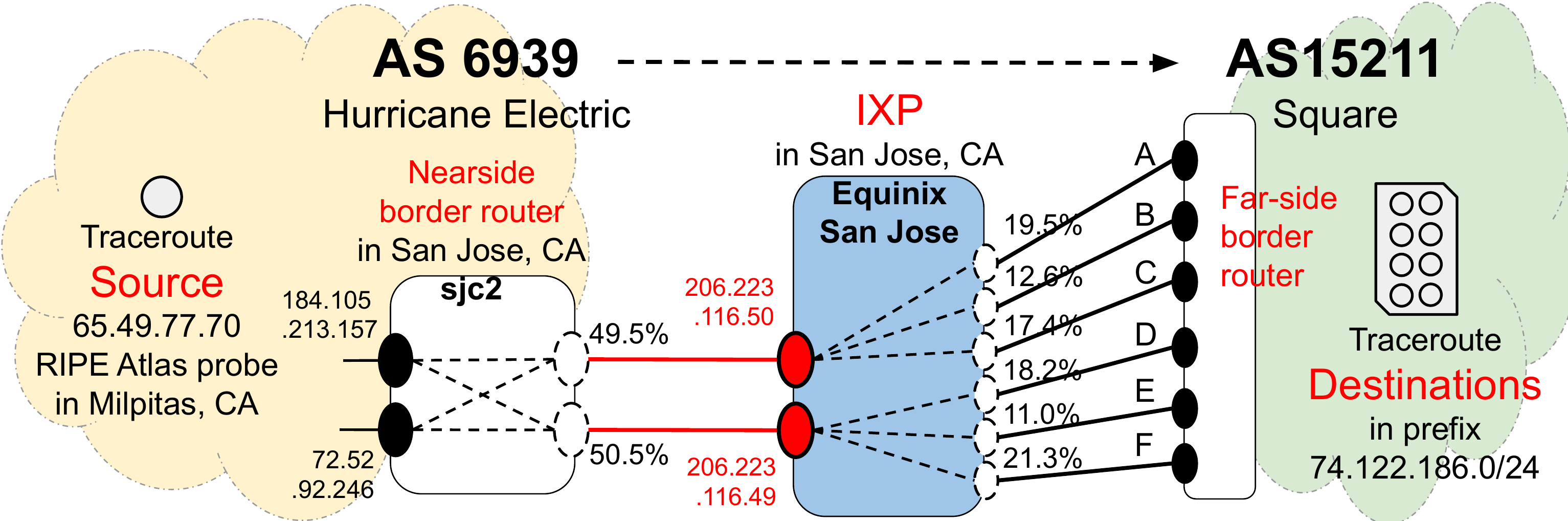}
    }
    \subfigure[Routing]{
    \label{fig:traffic_case3}
    \includegraphics[width = 0.36 \textwidth]{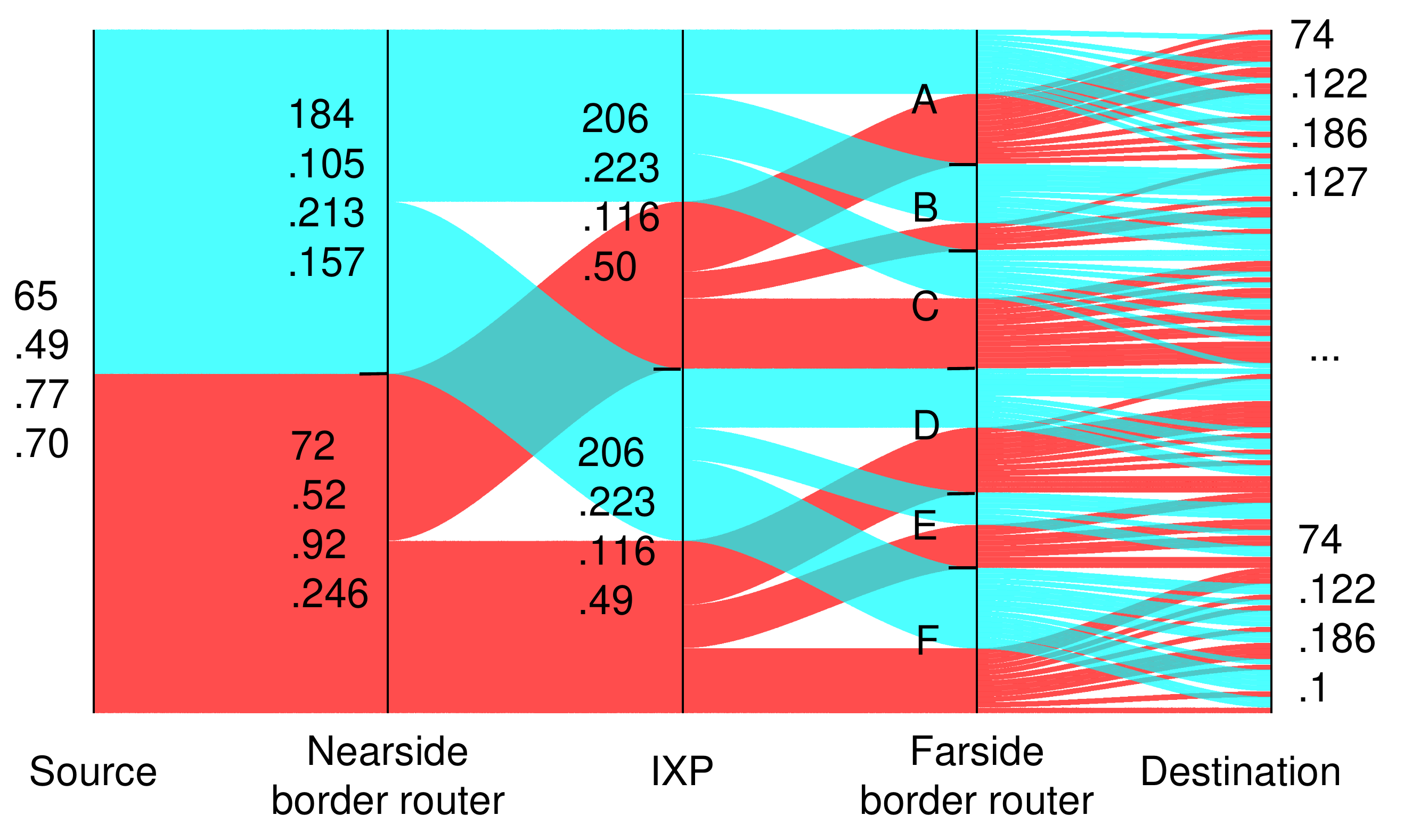}
    }
    
    \caption{Illustrations of topology and routing of a Divergent-type M-BGP deployment between Hurricane Electric and Square (Case 3 in Table \ref{tab:cases})
    }   
    \label{fig:divergent}
\end{figure*}

\section{Traceroute Measurement of M-BGP Deployment in Hurricane Electric}
\label{Sec:Traceroute}

This section introduces our traceroute measurement for revealing more details on Hurricane Electric's deployment of M-BGP. 

\subsection{RIPE Atlas traceroute measurement}
\label{Sec:Traceroute-RIPE}

Among the existing traceroute data or projects (e.g., RIPE Atlas, CAIDA Archipelago (Ark) \cite{Ark} and iPlane \cite{Madhyastha2006OSDI}), we use RIPE Atlas for our traceroute measurement. RIPE Atlas has deployed probes within Hurricane Electric, which enables us to probe from Hurricane Electric and ensures the traceroute paths will traverse the border routers. 

At the time we started the measurement, Hurricane Electric had three actively connected RIPE Atlas probes (IPv4) located in the United States, Canada and Iran. Because of the hot-potato rule, we expected traceroute sources to be geographically close to border routers with M-BGP deployment. We chose the probes in the United States (in Milpitas, CA, near border router {\tt sjc2} in San Jose, CA) and Canada (in Hamilton, near border router {\tt tor1} in Toronto). We did not use the probe in Iran because it is geographically far away from any border routers with M-BGP deployment. 

The destination prefixes are those identified with M-BGP deployment in the peering ASes to {\tt sjc2} and {\tt tor1}. We ran traceroute to each IP address  (from {\tt x.x.x.1} to {\tt x.x.x.254}) in the destination prefix. Each source-destination pair was probed 50 times with 7-minute interval in February 2020. We use RIPE Atlas default settings, namely ICMP and Paris traceroute~\cite{Augustin2006IMC}  variation 16.

\subsection{IP-to-AS mapping}
\label{Sec:Traceroute-IP2AS}

From traceroute raw data, we use IP-to-AS mapping  to locate border routers. There are many existing methods (e.g. bdrmapIT \cite{Marder2018IMC}, bdrmap \cite{Luckie2016IMC}, and MAP-IT \cite{Marder2016IMC}) and public datasets (e.g., MaxMind \cite{MaxMind} and Team Cymru \cite{Cymru}). 
As mentioned in Section \ref{Sec:LG}, these methods and datasets can be inaccurate on border identification. But here we do not use them to identify M-BGP; rather we use IP-to-AS mapping to analyse  traceroute data for M-BGP cases that we have already identified based on the ground-truth from LG queries. Also, to increase confidence,   we only accept result agreed by  both  bdrmapIT and RIPEstat  Data API~\cite{RIPEstat}.

bdrmapIT takes traceroute raw data as input, and integrates other datasets to conduct IP-to-AS mapping. The required datasets include prefixes announced by ASes (from RouteViews), IXP data (from PeeringDB \cite{PeeringDB}), customer cone data \cite{Luckie2013IMC}, AS relationship data \cite{caida_asrank} and AS-to-organization data \cite{AS2Org} (from CAIDA).  
The output of bdrmapIT is the AS that manages the router that an IP address (of an ingress interface) belongs to. In our traceroute data, bdrmapIT maps an IXP's IP to the AS of the next hop IP on the traceroute.

RIPEstat returns the AS that an IP belongs to. Normally, when an IP belongs to an IXP, RIPEstat does not provide an AS number for it. For example, according to RIPEstat, we obtain the IP-to-AS mapping result for an IP segment  $IP1-IP2-IP3$ as $IP1(AS1)-IP2(?)-IP3(AS2)$. Therefore, it is highly likely that IP2 belongs to an IXP. We rely on IXP data from PeeringDB and the following process for confirmation.
\begin{enumerate}[label=(\arabic*)]
    \item If IP2 belongs to a member AS of an IXP, it is mapped to the member AS (normally AS2); otherwise, go to (2). 
    
    \item If IP2 belongs in an IXP's own prefix, it is mapped to the AS of the next hop IP (in this example IP2 is mapped to AS2). Otherwise, the mapping is failed and this traceroute data is discarded. Note that in our study the traceroute is carried out between AS1 and AS2 (i.e., Hurricane Electric and one of its peering ASes), so it is impossible for IP2 to belong to a third member AS of the IXP. 
\end{enumerate}

Both bdrmapIT and RIPEstat can not map all IPs to ASes. In our study, they reach an agreement for 98\% of the overlapped IPs that both can successfully map. We only use the overlapped and agreed IP-to-AS mapping result for our analysis.  

If two consecutive IPs on a  traceroute path are mapped to different ASes, we use these IPs to represent (logically) an inter-domain border link, connecting between two border routers of two peering ASes, which are called nearside AS and farside AS.

\subsection{Two types of M-BGP deployment}
\label{Sec:Traceroute-TwoTypes}

All the M-BGP deployment studied in this paper are established through IXPs.  
Thus, a relevant traceroute path traverses firstly the nearside IP in Hurricane Electric, then an IP in an IXP  and finally a farside IP in the farside AS, where the IXP's IP is mapped to the farside AS as the result of IP-to-AS mapping used in this paper (see Section \ref{Sec:Traceroute-IP2AS}). This means the Next Hops or the Neighbor Addresses in the response of LG commands are actually IP addresses of interfaces of the IXP sitting between the nearside AS and the farside AS. 

Based on the traceroute data, we classify the identified M-BGP deployment into two types: parallel and divergent. 
Table \ref{tab:cases} lists the details of four cases, two in each type. Cases 1-3 are all from the same source IP (i.e., the same RIPE Atlas probe), hence the same border router and the same nearside IPs. Case 4 is from a different source IP (i.e., another RIPE Atlas probe), thus a different border router and different nearside IPs. We illustrate the topology and traffic for Cases 1 and 3 in Figures \ref{fig:parallel}-\ref{fig:divergent}, separately.

Table \ref{tab:cases} shows two cases of the parallel type M-BGP, in which each of the two IXP IPs is followed by a single farside IP. In Case 1, traffic enters the border router {\tt sjc2} via two nearside IPs; and then exits the border router  and enters a geographically nearby IXP (Equnix San Jose) via two IXP IPs with equal probabilities. The traffic from each IXP IP is forwarded to one of the two links  between IXP and farside AS. There is no cross traffic, i.e., there are only two unique paths between the nearside and the farside and traffic does not mix in the IXP.  

Figure \ref{fig:parallel} illustrates the topology   and routing of Case 1. The figure shows that the traffic is already split before entering {\tt sjc2}. We believe this is caused by intra-domain load sharing and {is independent} on M-BGP deployment because the traffic from either ingress interface of {\tt sjc2} is forwarded to the two IXP IPs equally, indicating a full mesh between ingress interfaces and egress interfaces of {\tt sjc2}.  

Figure \ref{fig:traffic_case1} also shows the traffic to each destination IP. This suggests two important observations. Firstly, each IXP IP is used for traffic to half of destination IPs. And secondly, the choice of IXP IP for each destination IP is permanent. That is, if an IXP IP is chosen for traffic to a particular destination IP, this IXP IP  will always be used for all future traffic to that destination. 
This is exactly the kind of routing property expected from M-BGP. 
The same  can be observed from the other cases.

Cases 3 and 4 in  Table \ref{tab:cases} are both divergent type, in which each IXP IP is followed by multiple farside IPs. We take Case 3 as an instance with its topology and routing shown in Figure \ref{fig:divergent}. In this case, traffic again exits {\tt sjc2} and enters Equinix San Jose via two IPs. Each IXP IP is used for traffic to 
half of IP addresses in the destination prefix. Traffic from each IXP IP is then split onto 3 different links between the IXP and the farside AS with similar 
proportions.

\section{Discussion}
\label{Sec:Discussion}
 
This paper reports our study on the deployment of   M-BGP in a large ISP network of Hurricane Electric or AS6939. We show that M-BGP is widely deployed by Hurricane Electric with hundreds of its peering ASes at more than half of its border routers around the globe. 
We   observe that most  of  its M-BGP  deployments  involve  IXP  interconnections.
All of our datasets are freely available at a GitHub repository \cite{GitHub}.

Since we only make queries to a limited number of prefixes that belong to each of Hurricane Electric's peering ASes,  
our result provides a lower bound of Hurricane Electric's deployment of M-BGP with its peering ASes. 
We focus on Hurricane Electric because of its very high centrality in the Internet routing system.
According to CAIDA's AS Rank~\cite{caida_asrank}, Hurricane Electric is the 7th largest AS in terms of customer cone size and provides transit between more than 8k ASes (12\% of ASNs in the global routing table).
Additionally, Hurricane Electric is a network of very high peering affinity, with more than 6k peers and presence at 236 IXPs (more than any other AS). 
Therefore, the extent of Hurricane Electric connectivity combined with its data transparency through the provided LG makes it an ideal vantage point for understanding the deployment of M-BGP in the Internet and evaluating the proposed measurement techniques. 

As part of our ongoing work, we are applying our technique to a much wider list of ASes hosting LG servers that support the required commands in order to provide a more extensive view of M-BGP deployments in the wild. Note that the M-BGP deployments presented in this paper are all via IXP and multiple inter-domain links. Our preliminary findings from a wider set of vantage points revealed M-BGP deployments using single inter-domain links or direct private peerings, and cases of multipath BGP routes with paths of unequal lengths. 
In addition we are expanding our traceroute measurements to evaluate the efficacy of MDA in discovering these M-BGP paths and to reveal potential non-canonical M-BGP deployments that use per-packet load balancing. 

We believe that the measurement, characterization and analysis of M-BGP is of particular interest to both network practitioners and Internet researchers.
The potential of M-BGP in improving the performance, stability and resilience of inter-domain paths, has not been yet thoroughly studied and understood.
Therefore, our work can inform and enable the necessary measurement studies to illuminate this crucial aspect of BGP.

\section*{Acknowledgment}
\label{Sec:Acknowledgment}

The authors would like to give special thanks to the anonymous reviewers for their constructing comments on the improvement of this paper.

%\vspace{12pt}
%\color{red}
%IEEE conference templates contain guidance text for composing and formatting conference papers. Please ensure that all template text is removed from your conference paper prior to submission to the conference. Failure to remove the template text from your paper may result in your paper not being published.

\end{document}